\documentclass{aa} 
\usepackage{graphicx}
\newcommand{\etal}{{et al. }}

\authorrunning{P. F. Chen et al.}
\titlerunning{Prominence Oscillation Before Eruption}

\begin{document}
   \title{SOHO/SUMER Observations of Prominence Oscillation Before Eruption}

   \author{P. F. Chen\inst{1}, D. E. Innes\inst{2}, and S. K. Solanki\inst{2}}

   \offprints{P. F. Chen}

\institute{Department of Astronomy, Nanjing University,
           Nanjing 210093, China\\
              \email{chenpf@nju.edu.cn}
         \and
           {Max-Planck-Institut f\"ur Sonnensystemforschung, 37191
      Katlenburg-Lindau, Germany \\
             \email{innes@mps.mpg.de; solanki@mps.mpg.de}}
             }

   \date{Received ......; accepted ......}


\abstract
   {Coronal mass ejections (CMEs), as a large-scale eruptive phenomenon,
    often reveal some precursors in the initiation phase, e.g., X-ray
    brightening, filament darkening, etc, which are useful for CME
    modeling and space weather forecast.
    }
   {With the SOHO/SUMER spectroscopic observations of the 2000 September
    26 event, we propose another precursor for CME eruptions, namely,
    long-time prominence oscillations.
    }
   {The prominence oscillation-and-eruption event was observed by
    ground-based H$\alpha$ telescopes and space-borne white-light, EUV
    imaging and spectroscopic instruments. In particular, the SUMER slit
    was observing the prominence in a sit-and-stare mode.
     }
   {The observations indicate that a siphon flow was moving from the
    proximity of the prominence to a site at a projected distance of
    270$''$, which was followed by repetitive H$\alpha$ surges and
    continual prominence oscillations. The oscillation lasted 4 hours
    before the prominence erupted as a blob-like CME. The analysis of
    the multiwavelength data indicates that the whole series of
    processes fits well into the emerging flux trigger mechanism for
    CMEs. In this mechanism, emerging magnetic flux drives a siphon
    flow due to increased gas pressure where the background polarity
    emerges. It also drives H$\alpha$ surges through magnetic
    reconnection where the opposite polarity emerges. The magnetic
    reconnection triggers the prominence oscillations, as well as its
    loss of equilibrium, which finally leads to the eruption of the
    prominence. It is also found that the reconnection between the
    emerging flux and the pre-existing magnetic loop proceeds in an
    intermittent, probably quasi-periodic, way.
}
   {}


   \maketitle

\section{Introduction}

Coronal mass ejections (CMEs) are large-scale phenomena on the Sun,
which may drive major magnetic storms on Earth, and hence have been
under extensive investigation in the context of space weather for
decades. CMEs often present thermal or
nonthermal signatures before their eruptions, which are called
precursors. The precursors of a CME are very useful since they can be
used to predict the occurrence of a CME, as well as to construct CME
initiation models (see Gopalswamy \etal \cite{gopal06} for a review).

The precursors found in the past decades can be summarised as follows:
(1) Emerging flux: Feynman \& Martin (\cite{feyn95}) found that two
thirds of the quiescent-filament-associated CMEs occurred after
substantial amounts of new magnetic flux emerged in the vicinity of the
filament, and that all filaments observed during a two-month period, in
which the new flux was oriented favourably for reconnection, erupted.
The merit of reconnection-favoured emerging flux as a precursor is
that it appears a few days before the CME eruption, so that the
eruption can be predicted a few days earlier. In order to
interpret such a high correlation, Chen \& Shibata (\cite{chen00})
proposed an emerging flux trigger mechanism for CMEs, where the
reconnection between the coronal field and the emerging flux,
either inside or outside the filament channel, restructures the
coronal magnetic configuration, leading to the loss of equilibrium
of the flux rope system. As a result, a current sheet forms below
the flux rope. The ensuing reconnection of the current sheet
results in the final eruption of the flux rope. Such a picture was
confirmed by Sterling \etal (\cite{ster05}; \cite{ster07}). (2)
X-ray brightening: Harrison \etal (\cite{harr85}) studied 6
CME/flare events and found that at a time coincident with the
projected onset of the CMEs, there is a small soft X-ray (SXR)
enhancement $\sim20$ min before the impulsive phase of the main
flare. Such an SXR enhancement might correspond to the
reconnection between the emerging flux and the pre-existing
magnetic field discussed above. (3) Radio noise storms: Lantos
\etal (\cite{lant81}) found that intense radio noise storms appear
before the eruption of a CME, which implies that the large-scale
magnetic structure evolves before the final disruption. Ramesh \&
Sundaram (\cite{rame01}), Chertok \etal (\cite{cher01}), and more
recently, Wen, Wang, \& Zhang (\cite{wen07}) verified that CMEs
are preceded by either the beginning or disappearance of noise
storms. (4) Type III radio burst groups: Jackson \etal
(\cite{jack78}) studied 40 CMEs observed by the Skylab
satellite and found that 5 to 10 hrs prior to the time when a CME
is first visible, the occurrence rate of type III radio bursts
increases by at least 2.5 times the average. The result might
also be understandable in the framework of the Chen \& Shibata
(\cite{chen00}) initiation model, where localised reconnection in
the trigger phase could accelerate electrons and produce type III
bursts well before the eruption of the CME. (5) Filament darkening
and widening: Martin (\cite{mart80}) summarised preflare
conditions and pointed out that filaments often become darker and
wider about 1 hr before the occurrence of a long duration flare,
which is the type that is always accompanied by a CME.

In this paper, we present extreme ultraviolet (EUV) spectroscopic
observations of  oscillations of a prominence before its eruption,
and propose that the long-time prominence oscillations are
another precursor of CMEs. Note that a filament is called a
prominence when it is located above the solar limb.

\section{Observations}

Two days after its first appearance, a prominence straddled over
the east limb on 2000 September 25, with the leading part sitting
on the solar disk as a filament. Starting from 20:50 UT on Sep. 25
and ending at 08:00 UT on Sep. 26, the slit of the Solar
Ultraviolet Measurements of Emitted Radiation (SUMER) spectrometer
on board the Solar and Heliospheric Observatory (SOHO) spacecraft
was positioned across the prominence, and spectral data were
recorded in the sit-and-stare mode. The 11 hr spectroscopic
observation covers the prominence eruption, as well as its
oscillation before the eruption. The dynamics of the prominence
prior to the eruption was also detected by the H$\alpha$ images
from Big Bear Solar Observatory (BBSO) and Mauna Loa Solar
Observatory (MLSO), while the activation and the ensuing eruption
were observed by the EUV Imaging Telescope (EIT; Delaboudini\`ere
\etal \cite{dela95}) with the \ion{Fe}{xii} 195 {\AA} filter and the
Large Angle and Spectrometric Coronagraph (LASCO; Brueckner \etal
\cite{brue95}) in white light, both being aboard the SOHO spacecraft.

The SUMER spectrometer (Wilhelm \etal \cite{wilh95}) is able to observe
simultaneously any selected 40 {\AA} window within its 660-1600 {\AA}
wavelength range. For the event analysed in this paper, emissions in the
window 1098-1138 {\AA} are obtained with a cadence of 169 s and a slit
of 4$''\times300''$. It contains both cold and hot lines in the
temperature range of $10^4$-$10^6$ K, including the chromospheric line
\ion{C}{i} 1118.45 {\AA} ($\sim 0.01$ MK), and transition region line
\ion{S}{iii}/\ion{Si}{iii} 1113 {\AA} ($\sim 0.05$ MK). The standard
procedure including calibrations and corrections is applied in the data
analysis.

BBSO recorded full-disk H$\alpha$ images with a high cadence (1 min)
and a pixel size of $\sim 1''$. Full-disk H$\alpha$ observations from
MLSO were also examined for reference.

\section{Results}

\begin{figure}
   \centering
   \includegraphics[width=15cm]{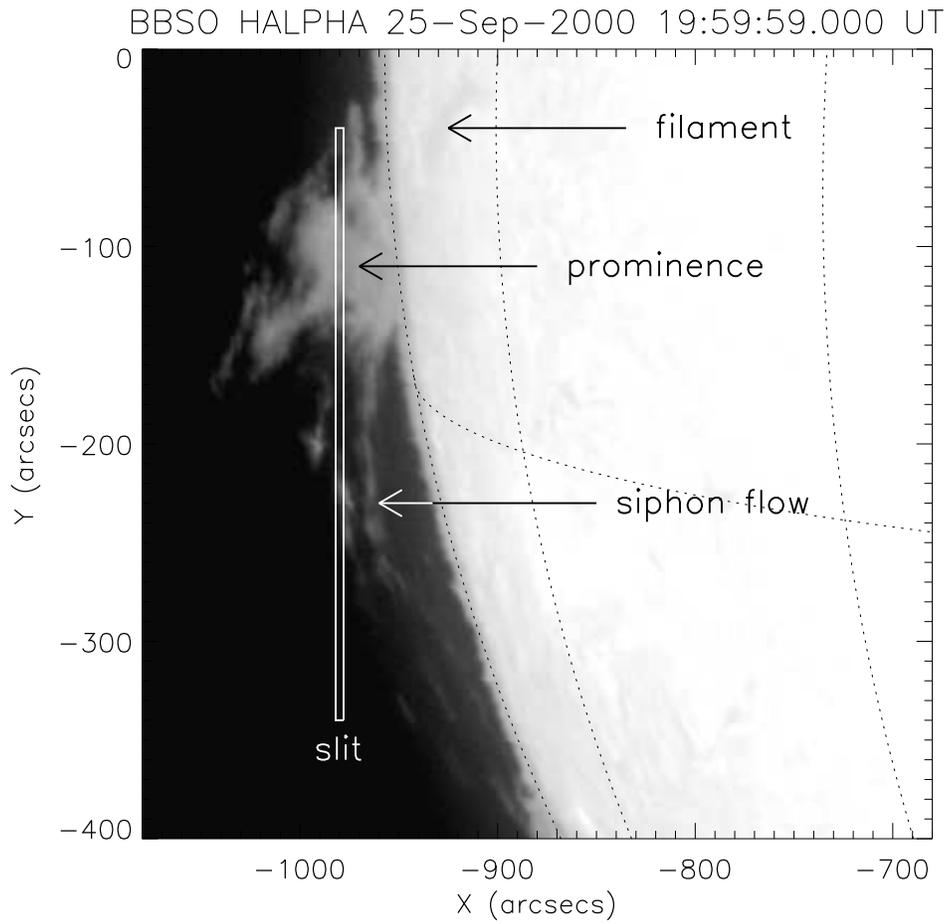}
      \caption{H$\alpha$ image from BBSO showing the positions of the
      prominence and the SUMER slit. Note that a siphon flow in moving
      to the south of the prominence.
    }
\label{fig1}
   \end{figure}

\begin{figure}
   \centering
   \includegraphics[width=18cm]{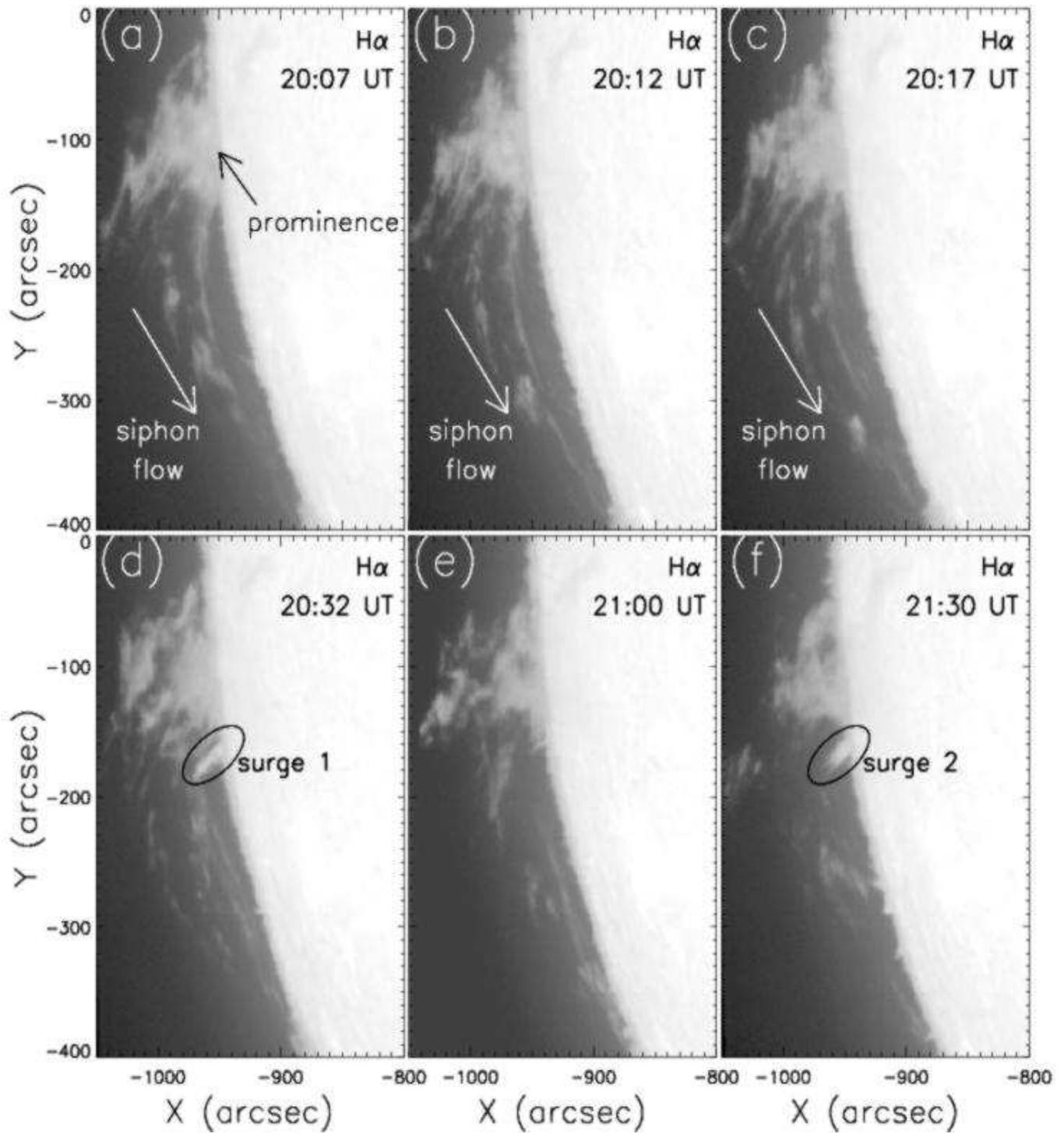}
      \caption{A time sequence of H$\alpha$ images, with the upper
      panels showing the siphon flow in a high cadence, and the
      lower panels showing the repetitive H$\alpha$ surges.}
\label{fig2}
   \end{figure}

The prominence, which was located above the solar east limb and $\sim
10^\circ$ to the south of the equator, extended to the northwest, where
it became visible as a faint filament on the disk as illustrated by the
H$\alpha$ image in Fig. \ref{fig1} (the filament is barely visible in
the figure as the brightness was adjusted to best show the details of
the prominence). The narrow rectangle in the figure indicates the
position of the SUMER slit. At least from 15:18 UT on Sep. 25 when the
first H$\alpha$ image in BBSO is available, a flow of chromospheric
material, presumably a siphon flow, was driven from the solar surface
in the proximity of the prominence along an arc pointing to the south
in the plane of the sky, probably guided by magnetic field lines.
Passing through the prominence in the projected plane, the
H$\alpha$-emitting plasma drained down to the solar surface at a place
about $270''$ to the south of the prominence, as shown in Fig.
\ref{fig2}. The figure displays the time sequence of the H$\alpha$
images, with a high cadence in the upper row to highlight the siphon
flow and a low cadence in the lower row to show the longer-term
dynamics. The pumping of the H$\alpha$ siphon flow from its origin
ended at $\sim$21:32 UT on Sep. 25. A time-slice analysis along the
trajectory of the flow indicates that the plasma was moving with a
speed of $\sim 60$ km s$^{-1}$, projected on the plane of the sky.

During the continual motion of the siphon flow, an H$\alpha$ surge was
seen to rise below the trajectory of the siphon flow at 20:27 UT, and
reached its peak at 20:32 UT, as indicated by the ellipse in panel (d)
of Fig. \ref{fig2}. The projected rise speed of the surge is estimated
to be $\sim 30$ km s$^{-1}$. The H$\alpha$ surge then fell down to the
chromosphere. About one hour later, i.e., at $\sim$21:26 UT, a second
surge erupted at the same site, as indicated by the ellipse in panel
(f). Another three surges erupted repetitively from the
same site at $\sim$22:26 UT, $\sim$22:42 UT, and $\sim$23:29 UT,
respectively.

\begin{figure}
   \centering
   \includegraphics[width=18cm]{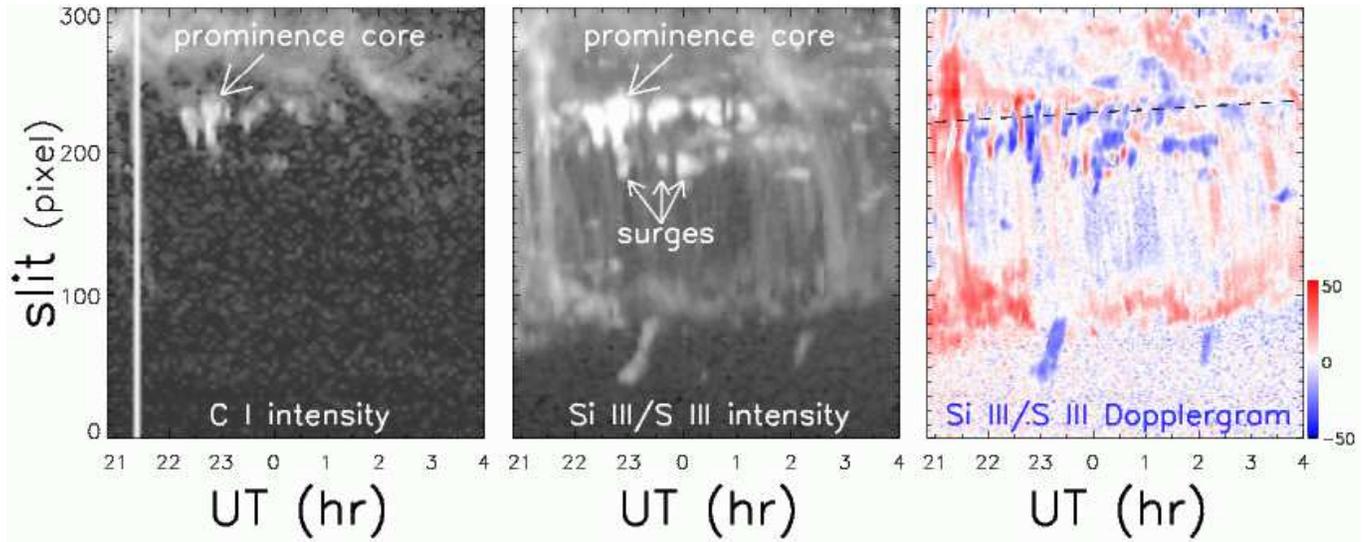}
      \caption{Left panel: Evolution of the \ion{C}{i} 1118.45 {\AA}
      intensity along the SUMER slit; Middle panel: Same for
      \ion{S}{iii}/\ion{Si}{iii} 1113 {\AA}; Right panel: Evolution of
      the Dopplergram along the the SUMER slit observed at
      \ion{S}{iii}/\ion{Si}{iii} 1113 {\AA}. The velocity evolution
      along the dashed cut line in the right panel is analysed in Fig.
      \ref{fig4}.}
\label{fig3}
   \end{figure}

The spectroscopic observation of SUMER started at 20:50 UT on Sep. 25.
The SUMER slit crossed the prominence, and coincidently, its lower
part covered the trajectory of the H$\alpha$ siphon flow, as seen in
Fig. \ref{fig1}. In order to see the prominence with the SUMER data,
the time evolution of the intensity of the cold line \ion{C}{i} 1118.45
{\AA} is plotted in the left panel of Fig. \ref{fig3}, where the core
of the prominence is manifest as bright knots around the 225th pixel
from the bottom of the slit. The temporal evolution of the line
intensity along the slit for the warmer line, \ion{S}{iii}/\ion{Si}{iii}
1113 {\AA}, is presented in the middle panel, and the corresponding
Dopplergram evolution is plotted in the right panel. The siphon flow is
also discerned in the intensity maps (clearly in the middle panel and 
slightly in the left panel of Fig. \ref{fig3}), as well as the
\ion{S}{iii}/\ion{Si}{iii} Dopplergram indicated by the nearly vertical
streaks, moving from the upper part to the lower part of the slit. Three
surges were detected by the SUMER slit, as indicated in the middle panel.
Comparison with the Dopplergram reveals that the surges initially
displayed a red shift, which then turned to blue shift.

The predominant red shift of the siphon flow implies that the strong
cool flow was moving away from the observer until 21:30 UT, with a line
of sight velocity of about 24 km s$^{-1}$. At 21:30 UT, the localised
prominence, as well as the large-scale siphon flow (seen in the
\ion{S}{iii}/\ion{Si}{iii} map), started to oscillate, as revealed by
the alternation of the red and blue Doppler shifts. The oscillation of
the prominence continued until 01:25 UT on Sep. 26. The oscillation
period is estimated to be $\sim20$ min by eye. In order to obtain the
oscillation period more precisely, we performed a wavelet analysis of
the Doppler velocity time series of the prominence core near the 225th
pixel from the bottom of the slit. Fig. \ref{fig4} shows the Doppler
velocity evolution in the left panel and its wavelet spectrum in the
right panel, which is obtained by the method described in Torrence \&
Compo (\cite{torr98}). Note that since the centroid of the prominence
was shifting slowly to the north, the velocity evolution in the left
panel of Fig. \ref{fig4} is taken along the dashed line in the right
panel of Fig. \ref{fig3}. The red regions indicate the locations of the
highest power, and the cross-hatched region corresponds to the cone of
influence. The wavelet spectrum confirms the 20 min period, and reveals
another longer period, around 60$\pm20$ min. After 01:25 UT, the
prominence disappeared from the slit, which is suggested by the absence
of bright knots in the left panel of Fig. \ref{fig3}. Note that Fig.
\ref{fig3} indicates that a weak \ion{S}{iii}/\ion{Si}{iii} siphon flow
persisted all the time throughout the event, contrary to the H$\alpha$
and \ion{C}{i} siphon flows, which were not visible after 21:32 UT on
Sep. 25. We also wish to mention that only a portion of the long
prominence erupted, and the rest stayed there even after the eruption.

In order to investigate the later evolution of the prominence, a time
sequence of the EUV 195 {\AA} images from SOHO/EIT and running
difference white-light images from the LASCO C2 coronagraph are
displayed in the upper and lower panels of Fig. \ref{fig5},
respectively. A careful examination of the EIT 195 {\AA} movie reveals
that, starting at least from 22:00 UT on Sep. 25, an elongated coronal
loop near the prominence was seen to be rising. At 01:25 on Sep. 26,
the coronal loop, embedding a void, expanded to a height about $250''$
above the solar limb (the coronal loop is outlined by the plus signs
and the void is marked by the arrow in the upper left panel of Fig.
\ref{fig5}). Subsequently, the prominence was lifted up, following
the expansion of the coronal loop.  At 02:00 UT (upper middle panel),
the prominence significantly deviated from its original position. At
03:48 UT (upper right panel), the prominence was out of the field of
view of SOHO/EIT, while flaring loops became visible below the expanding
coronal loop mentioned above after 02:36 UT. The flare was so weak that
it left no signature in the GOES 1-8 {\AA} light curve.

At 02:50 UT on Sep. 26, a bright feature was discerned in the
LASCO C2 field of view slightly south of the equatorial plane, as
shown by the lower left panel of Fig. \ref{fig5}. Later, a clearly
visible blob was seen to move almost radially, as indicated by the
white arrows in the lower middle and right panels. Owing to the
running difference method, a dark region immediately  follows the
bright blob. The bright blob looks like a typical CME core. In
this small CME event, no clear frontal loop was seen, probably due to
the limited instrumental sensitivity.

The heliocentric heights of the erupting prominence seen by SOHO/EIT
(only one data point) and the erupting blob observed by SOHO/LASCO
at various times are plotted in Fig. \ref{fig6}. It is found that the
erupting velocity of the white-light blob was $\sim 216$ km s$^{-1}$,
and the EUV prominence was roughly aligned with the height-time
profile of the white-light blob, confirming that the erupting
blob was most likely the erupting prominence.

\section{Discussions}

\subsection{Global picture}

Based on the analysis of all available observational data, the
whole process is summarised in Fig. \ref{fig7}, and described as
follows: Starting earlier than 16:00 UT on Sep. 25, a strong
H$\alpha$ siphon flow from the proximity of the prominence to a
site 270$''$ to the south was sustained until 21:32 UT. During this
period, there was one surge eruption near the prominence at
$\sim$20:32 UT. A second surge erupted at the same site at
$\sim$21:26 UT, and almost simultaneously, the prominence began to
oscillate, with a period of $\sim 20$ min. At the same time, an
elongated coronal loop above the prominence began to rise. The
prominence oscillation continued for 4 hrs, during which time
repetitive surges erupted at the same site. Starting at
$\sim$01:25 UT on Sep. 26, the prominence was pulled up and
ejected as a blob-like CME. Below the erupting prominence, flaring
loops appeared near the solar surface after 02:36 UT.

The late evolution of the event, from the coronal loop expansion to the
eruption of the prominence, as well as the ensuing solar flare fits
well into the standard model for CME/flare events, i.e, owing to some
kind of instability or loss of equilibrium, a coronal magnetic loop
rises, and an embedded flux rope, if one is present, subsequently also
rises, as the forces holding it down are reduced. As a result,
a current sheet forms below the flux rope. Magnetic reconnection in the
current sheet leads to the formation of flaring loops below the
reconnection point and the eruption of the flux rope above the
reconnection point (see Shibata \cite{shib05} for a historic review).
However, in this event it reamins unclear how the initial loss of
equilibrium was triggered, an essential problem in CME research, since
this event occurred near the limb, where the quality of the local
magnetogram was extremely poor and multiwavelength observations of
the low corona suffer seriously from the foreshortening effect.
Besides, a part of the activity was occulted by the solar limb.
Nevertheless, some clues can be gleaned on the basis of previous
research. It has been established that a siphon flow is driven by
enhanced gas pressure at one end of a coronal loop (Meyer \& Schmidt
\cite{meye68}), whereas H$\alpha$ surges are generated by magnetic
reconnection between emerging flux and the large-scale coronal field
(Kurokawa \& Kawai \cite{kuro93}; Yokoyama \& Shibata \cite{yoko95}).
Therefore, we can construct the following self-consistent paradigm
for this event, which is illustrated in Fig. \ref{fig8}: Coronal
magnetic loops AB and CD are field lines straddling over the
prominence and helping to keep it stable. Magnetic
flux emerges just outside the filament channel with the polarity
orientation drawn in the upper panel. Its inner (left) leg has
opposite magnetic polarity from the background and can therefore
trigger reconnection. The other (right) leg has the same polarity as
the background. Therefore, the merging flux squeezes the plasma
against the magnetic loop AB, which enhances the plasma density, as
well as the gas pressure near point A. The increased gas pressure
would drive a siphon flow along the magnetic field line from point A
to point B. Such a process may commence as soon as the new flux
emerges above the surface.  Note that, since the long prominence was
straddling over the solar east limb, and the fresh magnetic flux may
have emerged behind the limb (this is supported by the fact that the
H$\alpha$ surges did not show any signature on the disk), the siphon
flow was apparently seen to be running slightly below the prominence
on the projected plane. Between the emerging flux and the field line
CD, a current sheet forms since the left leg of the emerging flux has
the magnetic polarity opposite to the background (Solanki \etal
\cite{sola03}). After some time, e.g., when the current density (or
the electron drift speed in the
electric current) exceeds a threshold, reconnection is triggered
at point ``X", which would drive an H$\alpha$ surge. After the
reconnection, the field line CD becomes C$'$D. As demonstrated by
Chen \& Shibata (\cite{chen00}), the reconnected field line C$'$D
will expand in response to the magnetic restructuring. However, in
their 2D numerical simulations, the emerging flux and the
pre-existing coronal loops are in the same plane, so there is no
kink perturbation perpendicular to the loop plane. If the emerging
flux is inclined to the CD loop plane as plotted in Fig.
\ref{fig8}, the newly reconnected loops obtain a strong kink
perturbation. In this way, the magnetic loop C$'$D would oscillate
in the direction perpendicular to the loop plane, as indicated by
the double-headed arrow in the lower panel of Fig. \ref{fig8}. The
oscillation of the field line would drag the prominence to
oscillate in the same way. As demonstrated by Chen \& Shibata
(\cite{chen00}), the expanding coronal loop C$'$D finally does not hold
it down so strongly any more, so the prominence rises. The following
formation and disruption of a current sheet are well described by the
standard model, which was explained at the beginning of this paragraph.

Therefore, it is conjectured that it is the emerging flux that
drives the siphon flow and the reconnection with the pre-existing
coronal magnetic field, which leads to the formation of the
H$\alpha$ surges, triggers the oscillation and later eruption of
the prominence. The fact that reconnection-favoured emerging flux
can trigger the filament/prominence eruption has been verified by
Feynman \& Martin (\cite{feyn95}, see also Innes \etal \cite{inne99})
and explained by Chen \& Shibata
(\cite{chen00}). The observations presented here emphasise that
the magnetic reconnection, which occurs between the emerging flux
and the pre-existing coronal field in the CME triggering process,
takes place intermittently, rather than continually, as implied by
the repetitive behaviour of the H$\alpha$ surges. Such a
repetitive behaviour of H$\alpha$ surges, with a period of $\sim$1
hr, was reported by Schmieder \etal (\cite{schm84}). The
repetitive reconnection might be modulated by MHD waves in a way
similar to that proposed by Chen \& Priest (\cite{chen06}). The
repetitive reconnection can also account for the occurrence of the
type III radio burst group several hours before the CME as found
by Jackson \etal (\cite{jack78}).

\subsection{Prominence oscillations as a precursor of CMEs}

Prominence oscillations were first observed by Ramsey \& Smith
(\cite{rams66}). Since then, considerable progress has been made
due to both imaging and spectroscopic observations. Prominence
oscillations can be divided into large-amplitude type with
velocity amplitudes of the order of 20 km s$^{-1}$ (e.g., Ramsey
\& Smith \cite{rams66}) and small-amplitude type with velocity
amplitudes of the order of 2-3 km s$^{-1}$ (e.g., Thompson \&
Schmieder \cite{thom91}). Concerning the period ($P$), they can
also be categorised into three types (Molowny-Horas \etal
\cite{molo97}): short type ($P<$5 min, e.g., Thompson \& Schmieder
\cite{thom91}), intermediate type ($P\sim 6-20$ min, e.g.,
Bocchialini  \etal \cite{bocc01}), and long type ($P\sim$ 40 min
-- 2 hrs, e.g., Terradas \etal \cite{terr02}). After identifying
the oscillation mode, the properties of the oscillations,
including the period, damping time, wavelength, and so on, can
then be used to diagnose both thermal and magnetic parameters of
the prominence (see Roberts \cite{robe00}, Oliver \& Ballester
\cite{oliv02} for reviews).

Prominence oscillations can be caused by either an internal or
external agent (Vr\v{s}nak \cite{vrsn93}). The internal agent
might be some kind of instability of the prominence. For example,
Malville \& Schindler (\cite{malv81}) interpreted a
prominence oscillation in terms of kink instability. The external
agent might be Moreton waves from another active region (e.g.,
Ramsey \& Smith \cite{rams66}), photospheric 5-min and
chromospheric 3-min oscillations (e.g., Blanco \etal
\cite{blan99}), or photospheric convection (e.g., Malville
\cite{malv68}). The initiation of a CME is also a kind of
perturbation, which should trigger the oscillation of the
prominence/filament in the CME source region (note that almost
90\% of CMEs are associated with prominences, St Cyr \& Webb
\cite{stcy91}). For example, Malville \& Schindler (\cite{malv81})
were the first to find a prominence that oscillated for at least
90 min prior to the onset of a limb flare. From the theoretical
point of view, it is also reasonable to see prominence
oscillations before eruption. In the initiation models regarding
the photospheric shearing and convergent motions (e.g., Forbes \&
Priest \cite{forb95}), a kink instability would be responsible for
the triggering of the prominence oscillations; in the tether-cutting
initiation model (Moore \& Labonte \cite{moor80}), the first-step
reconnection of the internal magnetic field would induce a strong
perturbation to the prominence. Accordingly, in the emerging flux
trigger mechanism (Chen \& Shibata \cite{chen00}), magnetic
reconnection between the emerging flux and the pre-existing
magnetic field would impose a strong kink perturbation to the
magnetic field around the prominence. The lateral kink, as
depicted in Fig. \ref{fig8} and analysed here, would drive
horizontal string-type oscillations of the coronal magnetic field
and the filament. In the 2D numerical simulations of Chen \&
Shibata (\cite{chen00}), there was no $B_z$ in the magnetic
configuration. We have performed another simulation, with $B_z$
being introduced to the emerging flux. The preliminary results
indicate that the flux rope, which is believed to hold a filament
at the bottom, does show string-type oscillations in the direction
of the magnetic neutral line. Therefore, we expect to see
prominence oscillations in the initiation phase of a CME no matter
what the initiation mechanism is. That is to say, prominence
oscillations can be considered as another precursor of CMEs.

Since the moving direction of the string-mode oscillations is in the
horizontal plane and perpendicular to the filament thread, they are best
observed by spectroscopic instruments like SUMER when the filament is
above or near the solar limb and the filament thread is inclined to the
line of sight. Near the solar disk centre, such a string-mode
oscillation manifests itself as a spatial displacement. Consider that
the period is $\sim20$ min and that the velocity amplitude is $\sim10$
km s$^{-1}$, the amplitude of the displacement is $\sim1910$ km, or
 $\sim2.6$ arcsec, which could be resolved with high-resolution
imaging observations.

\begin{figure}
   \centering
   \includegraphics[width=18cm]{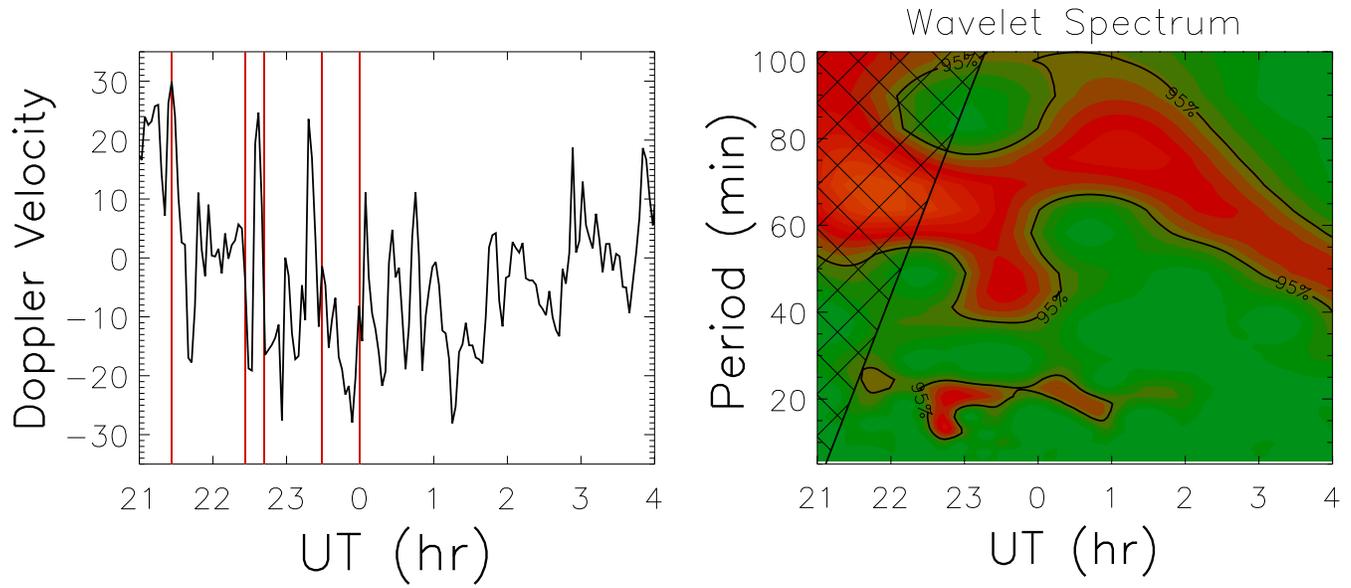}
      \caption{Time variation of the Doppler velocity of the prominence
	core (left panel) and its wavelet spectrum (right panel). In the
	left panel, the red vertical lines mark the occurrence of the
	surges; The right panel indicates that the prominence
	oscillations present a period at $\sim20$ min, as well as
	another one at $\sim60$ min.  }
\label{fig4}
   \end{figure}

\begin{figure}
   \centering
   \includegraphics[width=18cm]{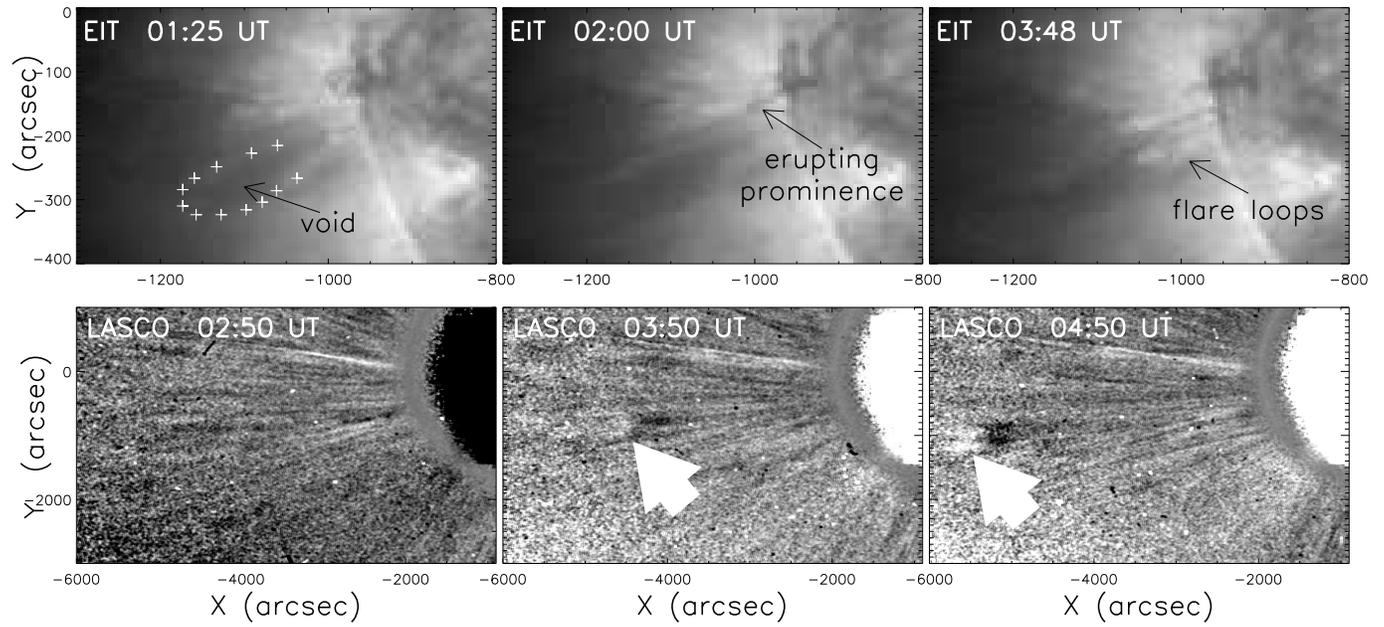}
      \caption{Upper panels: SOHO/EIT 195 {\AA} filter images showing
      the expansion of the coronal loop (outlined by the plus signs),
      filament eruption, and the flaring; Lower panels: SOHO/LASCO
      images showing the prominence eruption as a blob-like CME.}
\label{fig5}
   \end{figure}

\begin{figure}
   \centering
   \includegraphics[width=15cm]{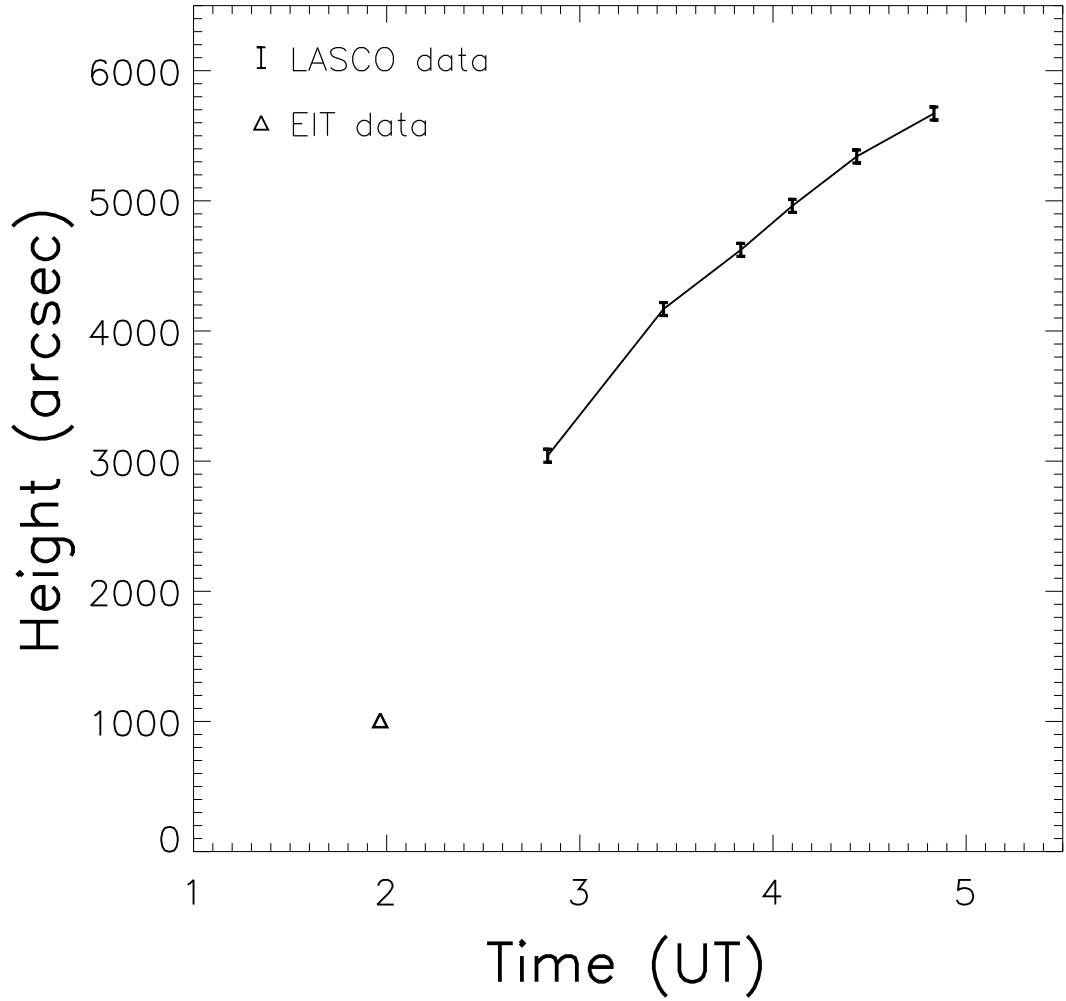}
      \caption{Height-time plot of the EUV prominence (triangle point)
      and the white-light blob CME (error bars connected by solid lines). }
\label{fig6}
   \end{figure}

\begin{figure}
   \centering
   \includegraphics[width=18cm]{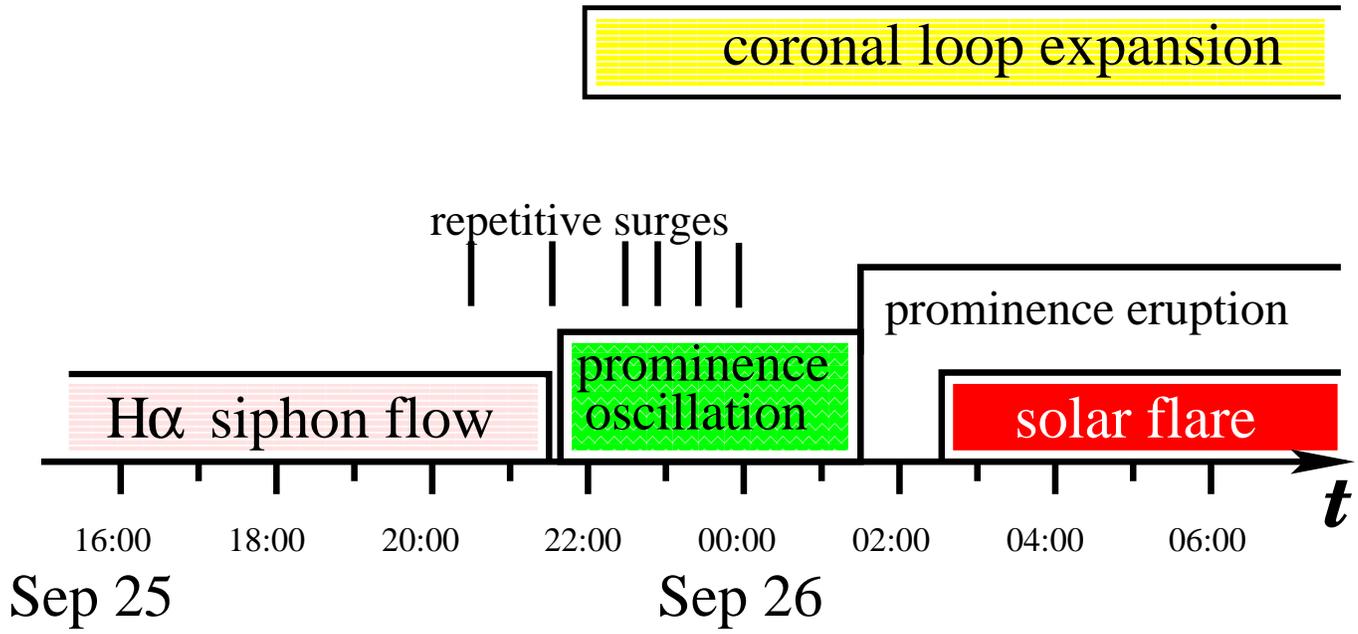}
      \caption{Timeline of all phenomena associated with the prominence
      eruption.}
\label{fig7}
   \end{figure}

\begin{figure}
   \centering
   \includegraphics[width=15cm]{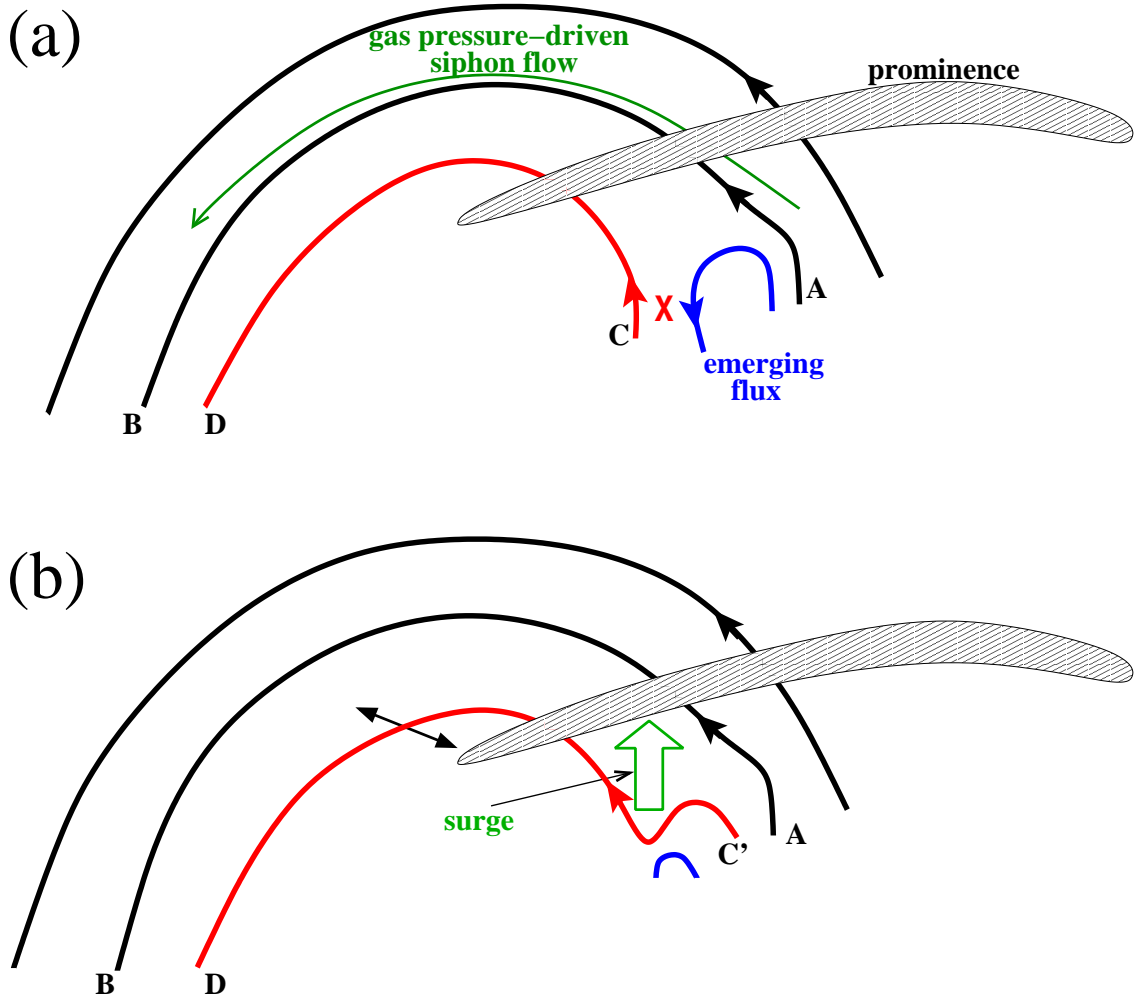}
      \caption{Sketch of the whole process, where the thick lines 
      represent magnetic field. Panel (a): emerging flux
      appears outside the filament channel, which drives the siphon
      flow near point A due to increased gas pressure and
      H$\alpha$ surges near point C due to magnetic reconnection;
      Panel (b): the kinked magnetic loop C$'$D after reconnection
      triggers the oscillations of the magnetic loops, along with the
      prominence, and the expansion of the coronal loops, which
      finally pulls the prominence to erupt.}
\label{fig8}
   \end{figure}

It is noted that prominences often oscillate, even in the
quiescent state (e.g., R{\'e}gnier, Solomon, \& Vial
\cite{regn01}), since perturbations  such as MHD waves from
sporadic eruptions, persistent $p$-mode oscillations and
convective motions in the photosphere, are ubiquitous in the solar
atmosphere.  No prominence eruption
follows these kinds of oscillations. However, for these
non-eruptive oscillations, the damping time is usually between 1
and 3 times the corresponding period (e.g., Tsubaki \& Takeuchi
\cite{tsub86}; Wiehr, Balthasar, \& Stellmacher \cite{wieh89};
Molowny-Horas \etal \cite{molo97}), whereas, our analysis of the
2000 September 26 event indicates that the prominence oscillations
lasted $\sim4$ hrs, almost 12 times the corresponding period,
before the final eruption. The reason is that the excitation
during the CME trigger phase is continuous, rather than just a
short pulse. The occurrence of the repetitive H$\alpha$ surges in
our event, at intervals of about 60 min, strongly suggests that the
reconnection between the emerging flux and the pre-existing
magnetic field proceeds in an intermittent, probably
quasi-periodic, way (see the timeline in Fig. \ref{fig7}). The
repetitive reconnection keeps shaking the magnetic loops around
the prominence, leading to the unusually long-time oscillations of
the prominence. This is supported by the almost one-to-one
correspondence of the H$\alpha$ surges and the occurrences of the
increased oscillation amplitude as shown in the left panel of Fig.
\ref{fig4}. Between the occurrence of the surges, the oscillation
amplitude did show fast damping.  
In this sense, different from the 20 min period of
the prominence oscillations, which is the intrinsic period of the
prominence, the 60 min period in the wavelet spectrum of the
prominence oscillation reflects the period of the repetitive
reconnection, which serves as an external driving agent for the
oscillations. Actually, the prominence oscillation can continue
even in the eruption phase, which was discovered by Isobe \&
Tripathi (\cite{isob06}). They noticed in their event that the
prominence oscillation was also caused by emerging flux. The results
of this work invite investigation of more prominences in order to
learn which fraction of them oscillate prior to eruption.

\begin{acknowledgements}
The authors thank MLSO and Haimin Wang at BBSO for providing H$\alpha$
data. Thanks are also due to H. Kurokawa for discussions about surges.
The research is supported by the Chinese foundations NCET-04-0445,
FANEDD (200226), 2006CB806302, and NSFC (10221001, 10333040 and
10403003). SOHO is a project of international cooperation between ESA
and NASA.

\end{acknowledgements}

\clearpage

\clearpage

\clearpage

\clearpage

\clearpage

\clearpage

\clearpage

\clearpage

\begin{thebibliography}{}

\bibitem[1999]{blan99} Blanco, S., Bocchialini,
K., Costa, A., Domenech, G., Rovira, M., \& Vial, J.-C.\ 1999, \solphys,
186, 281

\bibitem[2001]{bocc01} Bocchialini, K., Costa, A., Domenech, G., Rovira,
M., Vial, J.~C., \& Wingfield, K.\ 2001, \solphys, 199, 133

\bibitem[1995]{brue95} Brueckner, G.~E., et
al.\ 1995, \solphys, 162, 357

\bibitem[2006]{chen06} Chen, P.~F., \& Priest, E.~R.\ 2006, \solphys,
238, 313

\bibitem[2000]{chen00} Chen, P. F., \& Shibata, K. 2000, ApJ, 545, 524

\bibitem[2001]{cher01} Chertok, I.~M., Kahler,
S., Aurass, H., \& Gnezdilov, A.~A.\ 2001, \solphys, 202, 337

\bibitem[1995]{dela95}Delaboudini\'ere, J.-P. et al. 1995, \solphys,
162, 291

\bibitem[1995]{feyn95} Feynman, J., \& Martin, S.~F.\ 1995, \jgr, 100,
3355

\bibitem[1995]{forb95}Forbes, T. G. \& Priest, E. R. 1995,
      \apj, 446, 377

\bibitem[2006]{gopal06} Gopalswamy, N.,
Miki{\'c}, Z., Maia, D., Alexander, D., Cremades, H., Kaufmann, P.,
Tripathi, D., \& Wang, Y.-M.\ 2006, Space Science Reviews, 123, 303

\bibitem[1985]{harr85} Harrison, R.~A.,
Waggett, P.~W., Bentley, R.~D., Phillips, K.~J.~H., Bruner, M., Dryer, M.,
\& Simnett, G.~M.\ 1985, \solphys, 97, 387

\bibitem[1999]{inne99} Innes, D.~E., et al.\ 1999, \solphys, 186, 337 

\bibitem[2006]{isob06} Isobe, H., \& Tripathi, D.\ 2006, \aap, 449, L17

\bibitem[1978]{jack78} Jackson, B.~V.,
Sheridan, K.~V., Dulk, G.~A., \& McLean, D.~J.\ 1978, Proceedings of the
Astronomical Society of Australia, 3, 241

\bibitem[1993]{kuro93} Kurokawa, H., \& Kawai, G.\ 1993, IAU Colloq.
~141: The Magnetic and Velocity Fields of Solar Active Regions, 46, 507

\bibitem[1981]{lant81} Lantos, P., Kerdraon,
A., Rapley, G.~G., \& Bentley, R.~D.\ 1981, \aap, 101, 33

\bibitem[1968]{malv68} Malville, J.~M.\ 1968, \solphys, 4, 323

\bibitem[1981]{malv81} Malville, J.~M.,
\& Schindler, M.\ 1981, \solphys, 70, 115

\bibitem[1980]{mart80} Martin, S.~F.\ 1980, \solphys,
68, 217

\bibitem[1968]{meye68} Meyer, F., \& Schmidt, H.~U.\ 1968, Mitteilungen
der Astronomischen Gesellschaft Hamburg, 25, 194

\bibitem[1997]{molo97} Molowny-Horas,
R., Oliver, R., Ballester, J.~L., \& Baudin, F.\ 1997, \solphys, 172, 181

\bibitem[1980]{moor80} Moore, R.~L., \&
Labonte, B.~J.\ 1980, Solar and Interplanetary Dynamics, 91, 207

\bibitem[2002]{oliv02} Oliver, R., \&
Ballester, J.~L.\ 2002, \solphys, 206, 45

\bibitem[2001]{rame01} Ramesh, R., \&
Sundaram, G.~A.~S.\ 2001, \solphys, 202, 355

\bibitem[1966]{rams66} Ramsey, H.~E., \&
Smith, S.~F.\ 1966, \aj, 71, 197

\bibitem[2001]{regn01} R{\'e}gnier, S.,
Solomon, J., \& Vial, J.~C.\ 2001, \aap, 376, 292

\bibitem[2000]{robe00} Roberts, B.\ 2000, \solphys, 193, 139

\bibitem[1984]{schm84} Schmieder, B., Mein,
P., Martres, M.~J., \& Tandberg-Hanssen, E.\ 1984, \solphys, 94, 133

\bibitem[2005]{shib05} Shibata, K.\ 2005, IAU Symp., 226, 241 

\bibitem[2003]{sola03} Solanki, S.~K., Lagg, 
A., Woch, J., Krupp, N., \& Collados, M.\ 2003, \nat, 425, 692

\bibitem[1991]{stcy91} St.~Cyr, O.~C., \&
Webb, D.~F.\ 1991, \solphys, 136, 379

\bibitem[2007]{ster07} Sterling, A.~C.,
Harra, L.~K., \& Moore, R.~L.\ 2007, \apj, 669, 1359


\bibitem[2005]{ster05} Sterling, A.~C., \&
Moore, R.~L.\ 2005, \apj, 630, 1148

\bibitem[2002]{terr02} Terradas, J.,
Molowny-Horas, R., Wiehr, E., Balthasar, H., Oliver, R., \& Ballester,
J.~L.\ 2002, \aap, 393, 637

\bibitem[1991]{thom91} Thompson, W.~T.,
\& Schmieder, B.\ 1991, \aap, 243, 501

\bibitem[1998]{torr98} Torrence, C., \& Compo, G.P. 1998, Bull. Am.
	Meteorological Soc., 79, 61


\bibitem[1986]{tsub86} Tsubaki, T., \&
Takeuchi, A.\ 1986, \solphys, 104, 313

\bibitem[1993]{vrsn93} Vrsnak, B.\ 1993, Hvar Observatory Bulletin, 17,
23

\bibitem[2007]{wen07} Wen, Y.-Y., Wang, J.-X., \&
Zhang, Y.-Z.\ 2007, Chinese Journal of Astronomy and Astrophysics, 7, 265

\bibitem[1989]{wieh89} Wiehr, E., Balthasar, H.,
\& Stellmacher, G.\ 1989, Hvar Observatory Bulletin, 13, 131

\bibitem[1995]{wilh95} Wilhelm, K., et al.\ 1995, \solphys, 162, 189

\bibitem[1995]{yoko95} Yokoyama, T., \& Shibata, K.\ 1995, \nat, 375, 42

\end{thebibliography}
\end{document}